\documentstyle[12pt]{article}
\date{\today}

\title{ $ O(\alpha_s)$ QCD Corrections to Spin Correlations in $e^- e^+ \to
t \bar t $ process at the NLC }
\author{ Hong Xuan Liu$^1$, Chong Sheng Li$^1$,
Zhen Jun Xiao$^{1,2}$  \\
{\small 1 Department of Physics, Peking University, Beijing 100871,
China} \\
{\small 2 Department of Physics, Henan Normal University, Xinxiang,
Henan 453002, China.} \\ }
\date{\today}

\begin{document}
\maketitle
\begin{abstract}
Using a {\em Generic spin basis}, we present a general formalism of
one-loop radiative corrections to the spin correlations in the top
quark pair production at the Next Linear Collider,  and calculate
the $O(\alpha_s)$ QCD corrections under the soft gluon
approximation. We find that: (a) in {\em Off-diagonal basis}, the
$O(\alpha_s)$ QCD corrections to $e_L^- e^+$ ( $e_R^- e^+$ )
scattering process increase the  differential cross sections of the
dominant spin component $t_{\uparrow}\bar{t}_{\downarrow}$
($t_{\downarrow}\bar{t}_{\uparrow}$)  by $\sim 30\%$ and $\sim
(0.1\%-3\%)$ depending on the scattering angle for $\sqrt{s}=400
GeV$ and $1 TeV$, respectively; (b) in {Off-diagonal basis}
(Helicity basis), the dominant spin component makes up  $99.8\%$
($\sim 53\%$ ) of the total cross section at both tree and one-loop
level for $\sqrt{s}=400 GeV$, and the {\em Off-diagonal basis}
therefore remains to be the optimal spin basis after the inclusion
of $O(\alpha_s)$ QCD corrections.
\end{abstract}

\vspace{1.5cm}


\newcommand{\beq}{\begin{eqnarray}}
\newcommand{\eeq}{\end{eqnarray}}

\newcommand{\ssa}{\sin \theta_W}
\newcommand{\ssb}{\sin^2 \theta_W}
\newcommand{\cca}{\cos \theta_W}
\newcommand{\ccb}{\cos^2 \theta_W}

\newcommand{\tup}{t_{\uparrow}}
\newcommand{\td}{t_{\downarrow}}
\newcommand{\tbup}{\bar{t}_{\uparrow}}
\newcommand{\tbd}{\bar{t}_{\downarrow}}
\newcommand{\uu}{t_{\uparrow}\bar{t}_{\uparrow}}
\newcommand{\dd}{t_{\downarrow}\bar{t}_{\downarrow}}
\newcommand{\ud}{t_{\uparrow}\bar{t}_{\downarrow}}
\newcommand{\du}{t_{\downarrow}\bar{t}_{\uparrow}}

\newpage
A special feature of the top quark is that due to its large mass,
$m_t=175 \pm 6 GeV$\cite{top98}, the lifetime of the top quark is
very short. This has an important consequence that the top quark
decays before it hadronizes\cite{toph}, and the spin information is
preserved from production to decay. Thus we can expect the spin
orientation of the top quark to be observable experimentally. Since
there are significant angular correlations between the decay
products of the top quark and the spin of the top quark, the
spin-spin correlations in the top quark pair production can be
extracted by forming angular correlations among the decay products
of the top quark and the top anti-quark. Thus the spin correlations
in the top quark pair production can be used as a good observable
for testing the standard model (SM).

In most papers on the spin correlations for top quark pair production at
the Tevatron\cite{tt1} and the Next Linear Collider (NLC) \cite{tt2}, the
top quark spin is decomposed in the {\em Helicity basis}.
Recently, {\em Mahlon and Parke }\cite{parke96} proposed the {\em Generic
spin basis}  and found that the "Off-diagonal" basis, a special case of the
{\em Generic spin basis}, is a more optimized decomposition of the top
quark spins for $e^+ e^-$ colliders because the contribution from like-spin
pairs of top quarks vanishes to leading order  in perturbation theory.

The $O(\alpha_s)$ QCD corrections to the cross sections of $e^- e^+ \to
t \bar{t}$ process were calculated in refs.\cite{cross}.
In this letter, we first repeated the calculation of $O(\alpha_s)$
QCD corrections to this process.   We then present, in the {\em Generic
spin basis}, a general formalism
of one-loop radiative corrections to the spin correlations in the top quark
pair production at the NLC,  and calculate the $O(\alpha_s)$ QCD
corrections and give the numerical results.

In the SM, we consider the process
\beq
e^- e^+ \to t \bar t  \label{eett}
\eeq
at the NLC with $\sqrt{s}= (0.4, 1)$ TeV.
The tree level $Ve^-e^+$ vertex can be written as
\beq
\Gamma^\mu_{Vee} = ie\gamma^\mu(e^{V}_L P_- + e^{V}_R P_+)
\label{vee}
\eeq
where $P_{\pm}=(1\pm \gamma_5)/2$, and the SM values for these coupling
factors are  $e_{L,R}^\gamma=-1$ for $V=\gamma$, $e_L^Z=(2\ssb-1)/(2\ssa
\cca)$ and  $e_R^Z =\ssa/\cca$ for $V=Z$,  and the $\theta_W$ is the Weinberg
angle.

The general $Vt\bar t$ ($V=\gamma, Z$) coupling can be written as
\beq
\Gamma_{V t\bar{t}}^{\mu}& = & ie\left [ \gamma^\mu(A_V-B_V
\gamma^5)+\frac{t_1^\mu - t_2^\mu}{2}(C_V-D_V\gamma^5) \right ],
\label{vttb}
\eeq
where $t_1^\mu$ ($t_2^\mu$) is the momentum of the outgoing top (anti-top)
quark.

In the $t\bar t$ center of mass frame (CMS), the scattering plane is defined
to be the X-Z plane where the electron is moving along the $+Z$ direction and
$\theta_t$ is the scattering angle of top quark, and we also set $\phi_t=0$.
The Born helicity amplitudes for the process (\ref{eett}) are obtained
by summing the contributions from both the $Z$ and $\gamma$:
\beq
M_0(h_{e^-}, h_{e^+}, h_{t}, h_{\bar{t}})=
\frac{2e^2}{\sqrt{s}} \left [ M_0(h_{e^-}, h_{e^+}, h_{t},
h_{\bar{t}})^\gamma
+ M_0(h_{e^-}, h_{e^+}, h_{t}, h_{\bar{t}})^Z R(s) \right ].
\label{aht}
\eeq
where $s=4E^2$ is the total energy in CMS, $E$ is the energy of the
electron beam,  and $R(s)= s/(s-M_Z^2)$
\footnote{At NLC with $\sqrt{s} \geq 400 GeV$, the
imaginary part of the $Z$ propagator can be neglected safely.}.

In the {\em Generic spin basis}, as illustrated by Fig.1 in
ref.\cite{parke96}, the top quark (anti-top quark) spin states are
defined in the top quark (anti-top quark) rest-frame, where one
decomposes the top (anti-top) spin along the direction $\hat{s}_t$
($\hat{s}_{\bar t}$), which makes an angle $\xi$ with the anti-top
(top) momentum in the clockwise direction. Thus, the state $\tup
\tbup$ ($\td \tbd$) refers to a top with spin in the $+\hat{s}_t$
($-\hat{s}_t$) direction in the top rest-frame, and an anti-top
with spin $+\hat{s}_{\bar t}$  ($-\hat{s}_{\bar t}$ ) in the
anti-top rest-frame.

In the {\em Generic spin basis}, the amplitudes $M_0(h_{e^-}, h_{e^+},
\hat{s_{t}}, \hat{s_{\bar{t}}})$ for the process $e^- e^+ \to t \bar t$ can be generally
written as
\beq
M_0(-+\uu\; or \; \dd)& = & \mp \left[ m_tA_L\sin{\theta}\cos{\xi}-
(EA_L\cos{\theta}+KB_L)\sin{\xi} \right],\label{mg1}\\ M_0(-+\ud\:
or \; \du)& = &-(EA_L\cos{\theta}+KB_L)\cos{\xi}-
m_tA_L\sin{\theta}\sin{\xi}\nonumber\\ & &\mp
(EA_L+KB_L\cos{\theta}) ,\label{mg2} \\ M_0(+-\uu\; or \; \dd)& = &
\mp \left[ m_tA_R\sin{\theta}\cos{\xi}-
(EA_R\cos{\theta}-KB_R)\sin{\xi} \right],\label{mg3}\\ M_0(+-\ud\:
or \; \du)& = &-(EA_R\cos{\theta}-KB_R)\cos{\xi}
- m_tA_R\sin{\theta}\sin{\xi}\nonumber\\
& &\pm (EA_R-KB_R\cos{\theta}) ,\label{mg4}
\eeq
where $K = (E^2 -m_t^2)^{1/2}$. The amplitudes in {\em Helicity basis} can be
obtained easily by setting $\cos \xi =\pm 1$ in Eqs.(\ref{mg1})-(\ref{mg4}).
The form factors $A_{L,R}$ and $B_{L,R}$ are defined as
\beq
A_{L,R}& = &\frac{4e^2E}{s}\left ( e^{\gamma}_{L,R} A_\gamma
+ e^{Z}_{L,R} A_Z R(s)\right ), \label{alr}\\
B_{L,R}& =&\frac{4e^2E}{s}\left ( e^{\gamma}_{L,R} B_\gamma
+ e^{Z}_{L,R} B_Z R(s) \right ). \label{blr}
\eeq
At tree level,  the form factors
$(A_V, B_V)$ ($V=\gamma, Z$) appeared in
Eqs.(\ref{vttb}),(\ref{alr}) and (\ref{blr}) are
\beq
A_\gamma^0=\frac{2}{3}, \ \ B_\gamma^0=0,  \ \
A_Z^0=\frac{3-8\ssb}{12\ssa\cca}, \ \
B_Z^0=\frac{1}{4\ssa\cca}. \label{abcd}
\eeq

When we make a special choice for the angle $\xi$ in the
{\em Generic spin basis},
\beq
\tan{\xi} =
\frac{m_t A_L^0\sin{\theta}}{EA_L^0 \cos\theta + K B_L^0}\label{tan},
\eeq
the {\em Generic spin basis} turns into the so-called {\em Off-diagonal basis}
\cite{parke96} for $e_L^- e_R^+$ \\
scattering \footnote{As pointed in ref.\cite{parke96}, there are two {\em Off-diagonal
basis} for $e_L^- e_R^+$ and $e_R^- e_L^+$ scattering respectively. But these
two {\em Off-diagonal } bases are almost identical\cite{parke96}
and we also only use the Off-diagonal basis for $e_L^- e_R^+$ defined by
Eq.(\ref{tan}) even when discussing the case of $e_R^- e_L^+$ scattering.},
then we will have
\beq
\frac{d\sigma_0}{d\cos{\theta}}(- + \uu\; or \; \dd) &=& 0 \\
\frac{d\sigma_0}{d\cos{\theta}}(- + \ud\; or \; \du)
&=& (\frac{\beta}{32 \pi s})| M_0(- + \ud\; or \; \du)|^2\label{dcsa} \\
\frac{d\sigma_0}{d\cos{\theta}}(+ - \uu\; or \; \dd) &\approx & 0 \\
\frac{d\sigma_0}{d\cos{\theta}}(+ -  \ud\; or \; \du)
&=& (\frac{\beta}{32 \pi s})| M_0(+ -  \ud\; or \; \du)|^2
\label{dcsb}
\eeq
where the subscript $0$ means the tree level physical quantities.

At tree level, the form factors $A_V$ and $B_V$ ($V=\gamma, Z$) are
very simple, and $C_V=D_V=0$. When we consider the
one-loop corrections,  the form factors may become much more complicated.
In general, including the one-loop corrections, the form factors can be
written as
\beq
A_V=A_V^0 + \delta A_V, \ \ \ B_V=B_V^0 + \delta B_V
\label{ab}
\eeq
where $\delta A_V$ and $\delta B_V$ represent the one-loop corrections.

%%%%

At one-loop level, in the {\em Generic spin basis}, the general differential
cross sections for $e^- e^+$ scattering are
\beq
\frac{d\sigma}{d\cos{\theta}}(h_{e^-}, h_{e^+}, \hat{s}_t,
\hat{s}_{\bar t})= \frac{\beta}{32\pi s}  \left |
M_0(h_{e^-}, h_{e^+}, \hat{s}_t, \hat{s}_{\bar t} )
+ \delta M(h_{e^-}, h_{e^+}, \hat{s}_t, \hat{s}_{\bar t} )  \right | ^2,
\label{qcda}
\eeq
where $M_0$ is the tree level amplitudes given in
Eqs.(\ref{mg1})-(\ref{mg4}) and the $\delta M$ represents the
one-loop contributions, which are given by
\beq
\delta M(-+\uu or \dd) &=&
\mp \left[\rule[0mm]{0cm}{4mm} (m_t\delta A_L-K^2C_L)\sin{\theta} \cos \xi \right. \nonumber\\
&& - \left.  ( K\delta B_L + E\delta A_L\cos{\theta}) \sin \xi
\rule[0mm]{0cm}{4mm}\right ] - EK D_L \sin \theta , \\
\delta M(-+\ud or \du) &=& -(m_t\delta A_L-K^2C_L)\sin{\theta}\sin \xi
 - ( K\delta B_L + E\delta A_L\cos{\theta} ) \cos \xi \nonumber\\
&& \mp ( E\delta A_L + K\delta B_L\cos{\theta}) \\
\delta M(+-\uu or \dd) &=& \mp \left [ \rule[0mm]{0cm}{4mm}
(m_t\delta A_R-K^2C_R)\sin{\theta} \cos \xi \right. \nonumber\\
&& \left.  -  (  E\delta A_R\cos{\theta} - K\delta B_R ) \sin \xi
\rule[0mm]{0cm}{4mm}\right] - EK D_R \sin \theta  \\
\delta M(+-\ud or \du) &=& -(m_t\delta A_R-K^2C_R)\sin{\theta}\sin \xi
 - ( E\delta A_R\cos{\theta} - K\delta B_R  ) \cos \xi \nonumber\\
&& \mp (K\delta B_R\cos{\theta} - E\delta A_R  )
\eeq
with
\beq
\delta A_{L,R}& = &\frac{4e^2E}{s}\left ( e^{\gamma}_{L,R} \delta A_\gamma +
e^{Z}_{L,R} \delta A_Z R(s) \right ), \label{dalr}\\
\delta B_{L,R}& = &\frac{4e^2E}{s}\left ( e^{\gamma}_{L,R} \delta B_\gamma +
e^{Z}_{L,R} \delta B_Z R(s) \right ), \label{dblr}\\
C_{L,R}& =&\frac{4e^2E}{s}\left ( e^{\gamma}_{L,R} C_\gamma + e^{Z}_{L,R} C_Z
R(s)  \right ), \label{clr} \\
D_{L,R}&=& \frac{4e^2E}{s}\left ( e^{\gamma}_{L,R} D_\gamma +
e^{Z}_{L,R} D_Z R(s)  \right ), \label{dlr}
\eeq
The above formulae are valid for the general cases corresponding to the
vertex couplings as given in Eq.(\ref{vttb}).

In what follows, we will calculate the $O(\alpha_s)$ QCD corrections to
the top quark pair production as an interesting example. The relevant
Feynman diagrams include the QCD virtual corrections as well as the real
bremsstrahlung graphs.

We will use dimensional regularization to regulate
all the ultraviolet divergences in the virtual one-loop
corrections. To regulate the infrared divergences, we introduce a gluon mass
parameter $\lambda$ in the gluon propagator, which is justified since
the non-Abelian nature of QCD does not enter at the order in
$\alpha_s$. We also adopt the on-shell mass renormalization scheme.

For $O(\alpha_s) $ QCD virtual corrections, the form factor $D_V$
($V=\gamma, Z$) is still
zero,  while the non-zero  factors $\delta A_V$, $\delta B_V$ and $C_V$ are
\beq
\delta A_\gamma& = &\frac{\alpha_s}{3\pi}A_\gamma^0 U(\beta),\ \ \ \ \ \ \ \ \ \ \
\delta A_Z=\frac{\alpha_s}{3\pi} A_Z^0 U(\beta), \label{da1} \\
\delta B_\gamma &=&0, \ \ \ \ \ \ \ \ \ \ \ \ \ \ \ \ \ \ \ \ \ \ \
\delta B_Z=\frac{\alpha_s}{3\pi} B_Z^0 [ U(\beta)  -2 V(\beta)],
\label{db1}\\
C_\gamma& = &-\frac{\alpha_s}{3\pi m_t}C_\gamma^0 V(\beta),\ \ \ \ \ \
C_Z = -\frac{\alpha_s}{3\pi m_t}A_Z^0 V(\beta),\label{dc1}
\eeq
with
\beq
U(\beta)&=&\frac{1+\beta^2}{\beta}\left [
Sp(\frac{2\beta}{\beta-1})-Sp(\frac{2\beta}{1+\beta})+\pi^2 \right]-
3\beta\ln[ X_t ] -4  \nonumber\\
&& + \left [  \frac{ 1 + \beta^2}{\beta}\ln[ X_t]
+ 2 \right ] \ln[\frac{m_t^2}{\lambda^2}],
\eeq
where $\beta=\sqrt{1-4m_t^2/s}$,
$V(\beta)=(1-\beta^2)\ln[X_t]/\beta$, $X_t=(1-\beta)/(1+ \beta)$,
and $Sp(z)=-\int_{0}^{z}\ln{(1-t)}/t\,{\rm d}t$ is the Spence
function.  Since we have neglected the width of Z gauge boson,
only the real parts of $\delta A_V$, $\delta B_V$ and $C_V$ will
contribute to the corrections, and therefore we show only the real
parts of them in Eqs.(\ref{da1})-(\ref{dc1}).
In fact, the contributions from the imaginary parts of $\delta A_V$,
$\delta B_V$ and $C_V$ are negligibly small even if we keep the imaginary
parts of the $Z$ boson propagator, by our numerical calculations.

Although ultraviolet divergences have canceled in differential cross sections
in Eq.(\ref{qcda}), the infrared-divergent part is still present.

We regulate the infrared divergences associated with the soft
real-gluon emission using the same gluon mass parameter $\lambda$.
Under the soft gluon approximation (SGA), the amplitude induced by the
soft real gluon emissions can be written in a factorized form proportional to
the tree level amplitude. The corresponding differential cross sections therefore
are the form of
\beq
\frac{d\sigma_{soft}}{d\cos{\theta}}(h_{e^-}, h_{e^+}, \hat{s}_t,
\hat{s}_{\bar t})
= \eta_{soft}\frac{d\sigma_0}{d\cos{\theta}}
(h_{e^-}, h_{e^+}, \hat{s}_t, \hat{s}_{\bar t}) \label{dcssga}
\eeq
The factor $\eta_{soft}$ can be written as
\beq
\eta_{soft}=G_{IR}+G_{\delta}+G_{fin}
\eeq
Here $G_{IR}$ represents the infrared divergent part, $G_\delta$ is the
truncation part, and $G_{fin}$ is the finite part, and we have
\beq
G_{IR}& = &\frac{4\alpha _s}{3\pi}\ln[\frac{\lambda^2}{s}]
\left ( 1 + \frac{s-2m_t^2}{s\beta} \ln[X_t] \right ), \label{gir}\\
G_\delta& = &-\frac{4\alpha_s}{3\pi}\ln[\frac{4\omega_g^2}{s}]
\left ( 1+\frac{s-2m_t^2}{s\beta} \ln[X_t] \right ), \label{gdelta} \\
G_{fin}&=&-\frac{4\alpha_s}{3\pi}\left\{\frac{1}{\beta}\ln[X_t]
+\frac{s-2m_t^2}{s\beta}\left [ 2Sp(1-X_t)+\frac{1}{2}\ln
^2[X_t] \right ] \right\} \label{gfin}
\eeq
where  $\omega_g$ is the fixed maximum gluon energy.

All the infrared divergence will cancel after adding the real corrections
in Eq.(\ref{dcssga}) to the virtual corrections in Eq.(\ref{qcda}), as they must.
It is easy to see that the {\em Off-diagonal basis} defined at tree level
is still valid after the inclusion of QCD virtual corrections
and soft real gluon emission contributions. With the same value of $\xi$ as given in
Eq.(\ref{tan}), the $O(\alpha_s)$ QCD corrected differential cross sections
in the {\em Off-diagonal basis} are
\beq
\frac{d\sigma}{d\cos{\theta}}(-+\uu)&=&
\frac{d\sigma}{d\cos{\theta}}(-+\dd) =0  \\
\frac{d\sigma}{d\cos{\theta}}(-+\ud)&=&
\frac{d\sigma_0}{d\cos{\theta}}(-+\ud)\left\{ 1+\frac
{\alpha_s}{3\pi}\left [ X(\beta) + 2\left (
\frac{K V(\beta)}{|M_0(-+\ud)|}
\right. \right.  \right. \nonumber\\
&& \left. \left. \left.
\cdot ( \frac{KA_L}{m_t}\sin{\theta}\sin{\xi} - 2B_L\cos{\xi} -
2B_L\cos{\theta})\rule[0mm]{0cm}{6mm}\right ) \right ] \right\},  \\
\frac{d\sigma}{d\cos{\theta}}(-+\du)&=&
\frac{d\sigma_0}{d\cos{\theta}}(-+\du)\left\{ 1+\frac
{\alpha_s}{3\pi}\left [ X(\beta) + 2\left (
\frac{K V(\beta)}{|M_0(-+\du)|}
\right. \right.  \right. \nonumber\\
&& \left. \left. \left.
\cdot ( \frac{KA_L}{m_t}\sin{\theta}\sin{\xi} - 2B_L\cos{\xi}+
2B_L\cos{\theta})\rule[0mm]{0cm}{6mm}\right ) \right ] \right\}\\
\frac{d\sigma}{d\cos{\theta}}(+-\uu)&=&
\frac{d\sigma}{d\cos{\theta}}(+-\dd) \approx  0 \\
\frac{d\sigma}{d\cos{\theta}}(+-\ud)&=&
\frac{d\sigma_0}{d\cos{\theta}}(+-\ud)\left\{ 1+\frac
{\alpha_s}{3\pi}\left [ X(\beta) + 2\left ( \frac{K V(\beta)}{|M_0(+-\ud)|}
\right. \right.  \right. \nonumber\\
&& \left. \left. \left.
\cdot ( \frac{KA_R}{m_t}\sin{\theta}\sin{\xi} + 2B_R\cos{\xi} -
2B_R\cos{\theta})\rule[0mm]{0cm}{6mm}\right ) \right ] \right\},  \\
\frac{d\sigma}{d\cos{\theta}}(+-\du)&=&
\frac{d\sigma_0}{d\cos{\theta}}(+-\du)\left\{ 1+\frac
{\alpha_s}{3\pi}\left [ X(\beta) + 2\left (
\frac{K V(\beta)}{|M_0(+-\du)|}
\right. \right.  \right. \nonumber\\
&& \left. \left. \left.
\cdot ( \frac{KA_R}{m_t}\sin{\theta}\sin{\xi}+2B_R\cos{\xi}+
2B_R\cos{\theta})\rule[0mm]{0cm}{6mm}\right ) \right ] \right\}
\eeq
with
\beq
X(\beta) &=& \frac{1+\beta^2}{\beta}\left [
Sp(\frac{2\beta}{\beta-1})-Sp(\frac{2\beta}{1+\beta})+\pi^2 \right]
+ 2\left[ \frac{1+ \beta^2}{\beta}\ln[X_t]
+ 2 \right ] \ln{\frac{m_t^2}{4\omega_g^2}} \nonumber\\
&& -\frac{4(1+\beta^2)}{\beta}Sp(1-X_t) -\frac{4 +3\beta^2}{\beta}\ln[X_t]
- \frac{1+\beta^2}{\beta}\ln^2[X_t] - 4, \ \
\eeq
From above analytical expressions, we can see that the one-loop QCD
corrections do not change the spin configuration of top quark pairs.
For $e_L^- e_R^+$ scattering,  the contribution from the up-up (UU) and
down-down (DD) top quark pairs vanish at the $O(\alpha_s)$
order in {\em Off-diagonal spin basis}. For the $e_R^- e_L^+$ scattering,
the contributions from the UU and DD pairs of top quarks are also
very close to zero.
The cross sections in {\em Helicity basis} can be
obtained easily by setting $\cos \xi = 1$ in the above formulae.

In the numerical calculation, we use the following parameters as standard
input\cite{pdf98},
\beq
\alpha_s(M_Z)&=&0.118, \ \ m_Z=91.187 GeV, \ \
\sin^2{\theta_W}=0.2315,\nonumber\\
\alpha&=&\frac{1}{128}, \ \ m_t=175 GeV, \ \ \Gamma_Z=2.491 GeV.
\eeq
For the running of $\alpha_s$, we use $\sqrt{s}$ as the
renormalization scale. For the infra-red cutoff $\omega_g$ we take
$\omega_g =(\sqrt{s} -2m_t)/5$ \cite{9807209} under the soft gluon
approximation.

In the {\em Off-diagonal spin basis}, only one spin component is appreciably
non-zero:  the up-down $(\ud)$ component for $e_L^- e^+$ scattering and
the down-up $(\du)$ component for $e_R^- e^+$ scattering, as illustrated in
Fig.1. In the {\em Off-diagonal spin basis}, we have
\beq
\frac{d\sigma}{d\cos{\theta}}(e_L^- e^+ \to \ud)=\left \{ \begin{array}{ll}
(0.23 \sim 0.70) \ \ pb, & {\rm at \ \ tree \ \ level}, \\
(0.33 \sim 0.89)\ \  pb,& {\rm with \ \ QCD \ \ correction},
\end{array} \right. \label{dcsaa}
\eeq
and
\beq
\frac{d\sigma}{d\cos{\theta}}(e_R^- e^+ \to \du)=\left \{ \begin{array}{ll}
(0.09 \sim 0.33)\ \  pb, & {\rm at\ \ tree\ \ level}, \\
(0.16 \sim 0.43)\ \ pb,& {\rm with\ \  QCD \ \ correction},
 \end{array} \right. \label{dcsab}
\eeq
for $-1 \leq \cos \theta \leq 1$ and $\sqrt{s}=400 GeV$. Above
results show that, the QCD corrections make the  differential cross
sections of dominant spin components larger by $\sim 30\%$ compared
to the tree level ones for both $e_L^- e^+$ and $e_R^- e^+$ modes.
For $\sqrt{s}=1000$ GeV, however, the increase of the corresponding
differential cross sections induced by the inclusion of QCD
corrections is  less than $3\%$ as shown in Fig.2.

Fig.3 shows the fractions of the total cross sections for the dominant spin
components, defined in both the {Helicity basis} and the optimal {\em
Off-diagonal basis} for $e_L^- e^+$ scattering, as a function of the
total energy $\sqrt{s}$. The fractions were defined as
\beq
R_{UD} &=& \frac{\sigma_{UD}}{\sigma_{UD} + \sigma_{DU}},\\
R_{LR} &=& \frac{\sigma_{t_L \bar{t}_R}}{\sigma_{t_L \bar{t}_L} +
\sigma_{t_L \bar{t}_R} + \sigma_{t_R \bar{t}_L}+  \sigma_{t_R \bar{t}_R}}.
\eeq
In the {\em Off-diagonal spin basis}, the
dominant spin component $\ud$ makes up the $99.8\%$ ($96.4\%$) of the total
cross section for $\sqrt{s}=400 GeV$ ($1 TeV$) after including the
$O(\alpha_s)$ QCD corrections.
In the {\em Helicity basis}, however, the fraction for the dominant
spin component $t_L \bar{t}_R$ is only $53\%$ at $\sqrt{s}=400GeV$,
as shown by the lower two curves in Fig.4.
Although this fraction for $t_L \bar{t}_R$ will increase to $\sim 83\%$ at
$\sqrt{s}=1 TeV$, it is still less than the corresponding ratio $R_{\ud}
= 96.4\%$ at $\sqrt{s}=1 TeV$.
For the $e_R^- e^+$ scattering we have similar results.

When defined in the optimal {\em Off-diagonal basis}, the spins of
the $t$ and $\bar{t}$ pairs produced from polarized $e_L^- e^+$ and
$e_R^- e^+$ scattering are uniquely determined, and  this
specific  feature will not be changed after including the
$O(\alpha_s)$ QCD corrections under the soft-gluon approximation.
Because the contribution to spin correlations in the top quark pair
production from the hard-gluon emissions is numerically very small,
we here present only the results under the soft-gluon approximation.

For the sake of cross check, we also present the numerical results
in the ordinary {\em Helicity basis}.  Fig.4 shows the differential cross
sections for the process $e_L^- e^+ \to t_{L,R} \bar{t} $ and
$e_R^- e^+ \to t_{L,R} \bar{t} $ in the {\em Helicity basis } with
$\sqrt{s}=400GeV$, and the corresponding differential cross sections
are clearly well consistent with those shown in Fig.8 of
ref.\cite{9807209}.

To summarize, we have calculated the $O(\alpha_s)$ QCD corrections to the
spin correlations in the top quark pair production at NLC in
the SM. We start from the general forms of $Vt \bar{t}$
vertex ($V= \gamma, Z$), derive out the amplitudes in  the {\em Generic spin
basis}, and give the general formalism of including one-loop corrections
to the differential cross section in top quark pair production at the NLC,
and finally calculate the $O(\alpha_s)$ QCD corrections. We  found that:

\begin{quotation}

(a) In {\em Off-diagonal basis}, the $O(\alpha_s)$ QCD corrections
to $e_L^- e^+$ ( $e_R^- e^+$ ) scattering process increase the
differential cross sections of the dominant spin component
$t_{\uparrow}\bar{t}_{\downarrow}$
($t_{\downarrow}\bar{t}_{\uparrow}$)  by $\sim 30\%$ and
$\sim(0.1\%-3\%)$ depending on the scattering angle for
$\sqrt{s}=400 GeV$ and $1 TeV$, respectively.

(b) The {\em Off-diagonal basis} remains the optimal spin basis even after
the inclusion of $O(\alpha_s)$ QCD corrections. At $\sqrt{s}=400GeV$, the
dominant spin components in both $e_L^- e^+$ and $e_R^- e^+$  scattering
make up more than $99\%$ of the total cross section at both tree and one-loop
level, but such fraction is only $\sim 53\%$ in the {\em Helicity basis}.

\end{quotation}

\vspace{1cm}
 Note added. While preparing this manuscript the paper of J. Kodaira,
T. Nasuno and S. Parke appeared\cite{9807209} where the QCD corrections
to spin correlations in top quark production at $e^+e^-$ colliders
is also calculated. Their numerical results under the soft gluon
approximation are in very good agreement with ours.

\vspace{2.5cm}
\hspace*{4.5cm}ACKNOWLEDGMENTS\\

This work was supported in part by the National Natural Science Foundation
of China,  a grant from the State Commission of Science and technology
of China. The Project Supported by Doctoral Program Foundation of Institution
of Higher Education. Z.J. Xiao acknowledges the support by the National
Natural Science Foundation of China under the Grant No.19775012, and by
the funds from the Outstanding Young Teacher Foundation of the Education
Ministry of China.

\vspace{1cm}
%\newpage

\newpage

\begin{center}
{\bf Figure Captions}
\end{center}
\begin{description}

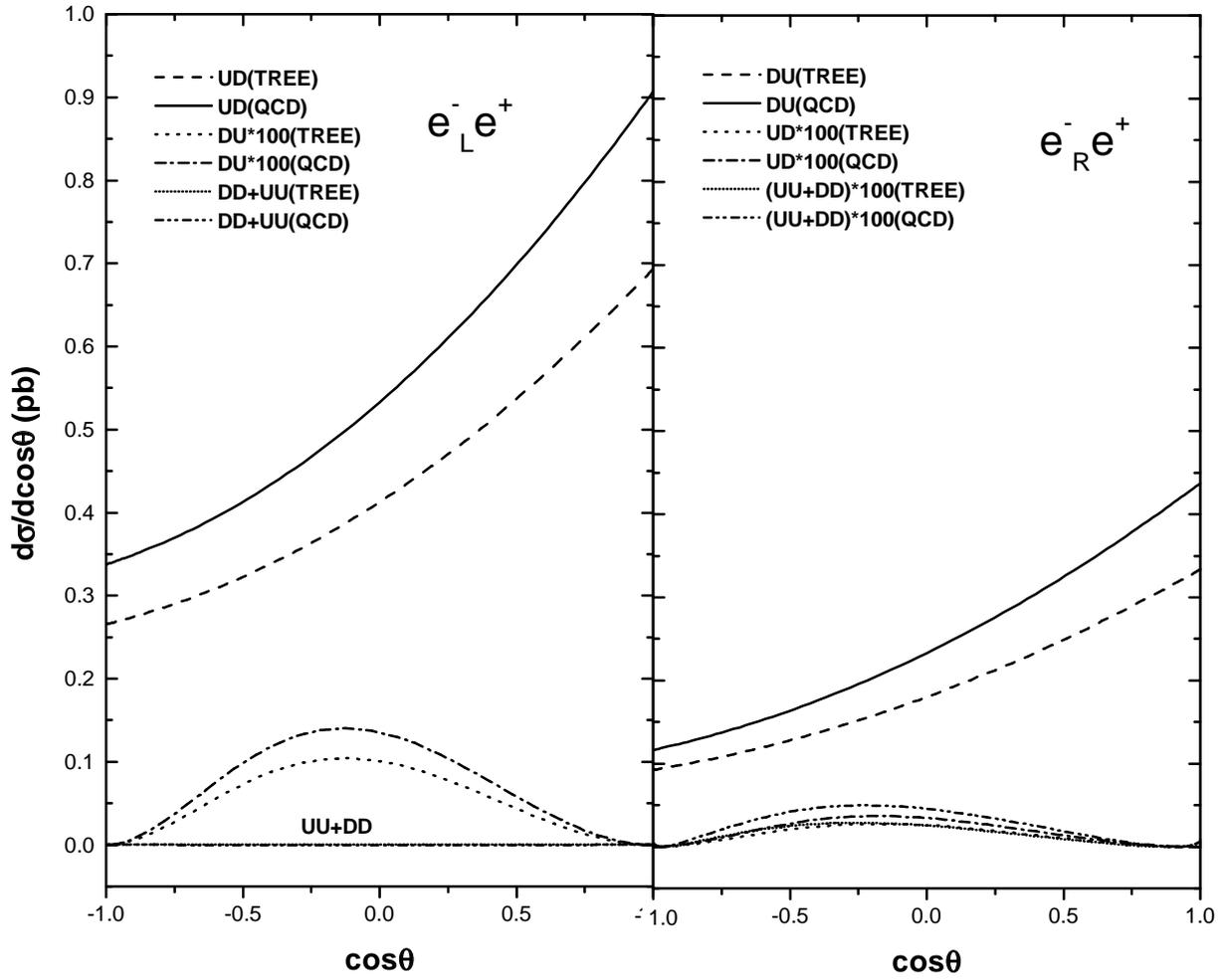
\item[Fig.1:] The differential  cross sections in the {\em Off-diagonal
basis} for the $e_{L,R}^- e^+ \to t\bar{t}$ processes UD ($\ud$),
DU ($\du$)  and  UU + DD ($\uu + \dd$), assuming $\sqrt{s}=400 GeV$
and $\omega_g=(\sqrt{s}-2m_t)/5=10 GeV$. The non-dominant $\du$ ($\ud$)
component for
the $e_L^-e^+$ ($e_R^- e^+$) scattering is multiplied by a factor of 100.

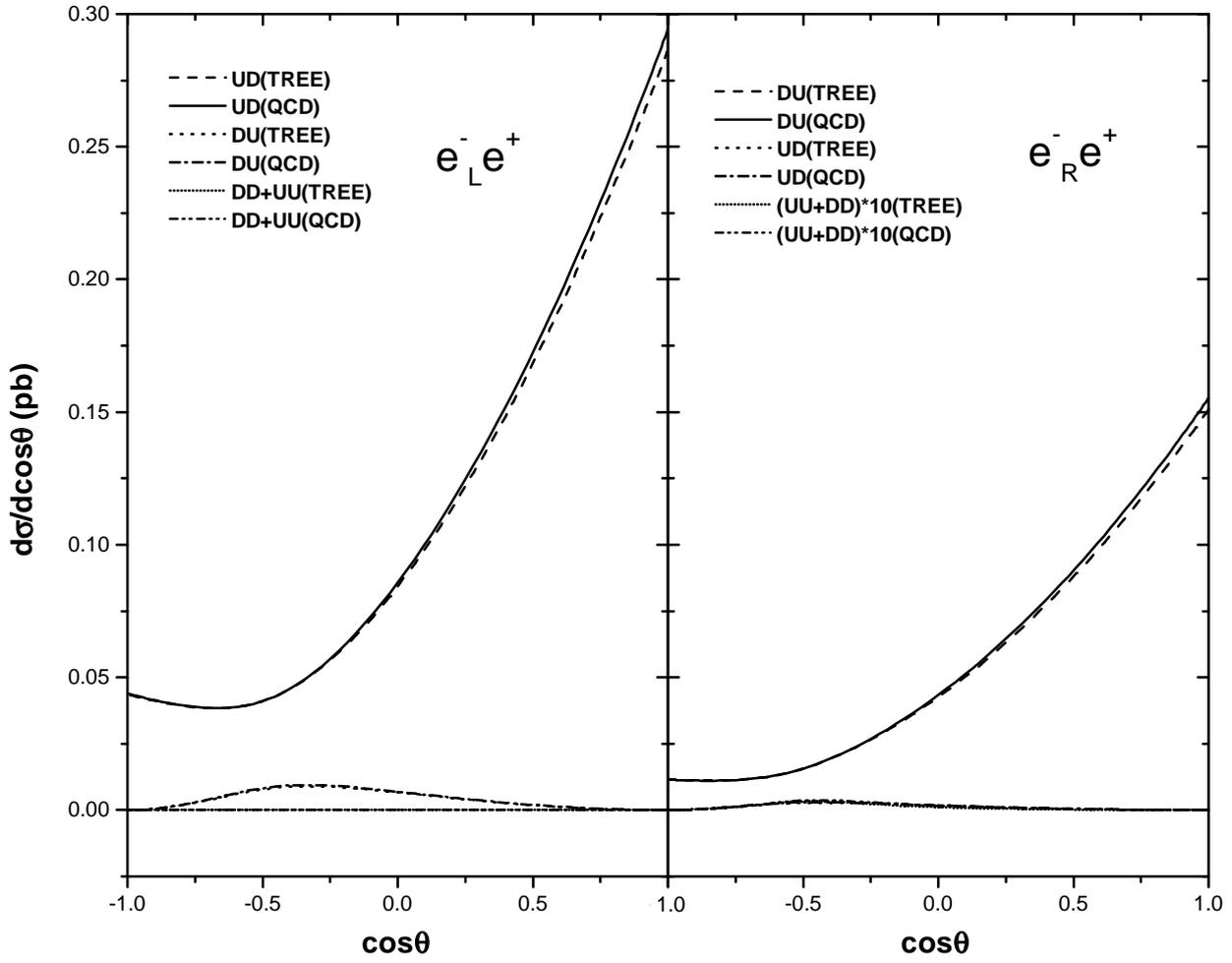
\item[Fig.2:] The same as Fig.1 but for $\sqrt{s}=1 TeV$ and
$\omega_g=(\sqrt{s}-2m_t)/5=130 GeV$. The non-dominant $\uu + \dd$ component
for the $e_R^- e^+$ scattering is multiplied by a factor of 10.

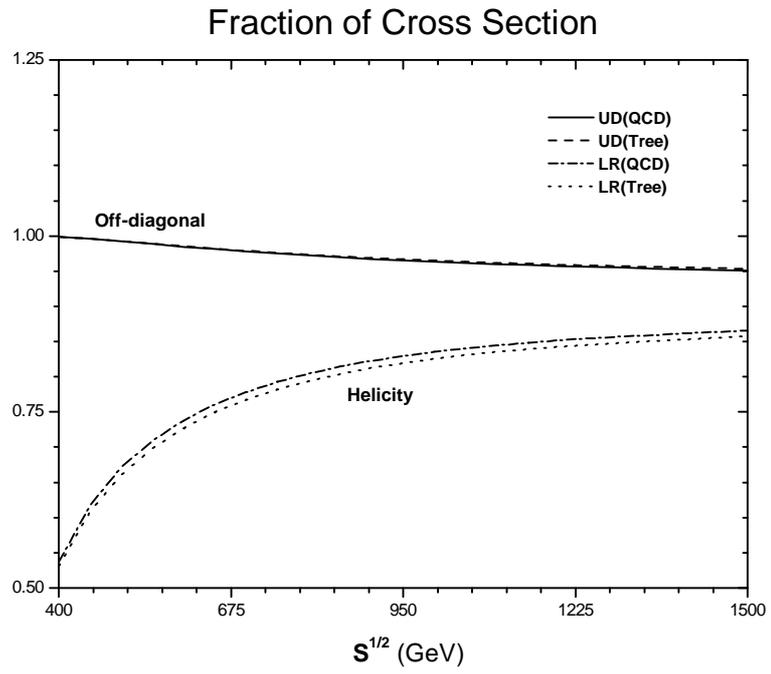
\item[Fig.3:] The $\sqrt{s}$ dependence of fractions of total cross sections
for dominant $\ud$ $(t_L \bar{t}_R)$ component for $e_L^- e^+$ scattering.
The two upper curves show the  fractions in the {\em Off-diagonal basis},
while the two lower curves correspond to the fractions in the
{\em Helicity basis}.

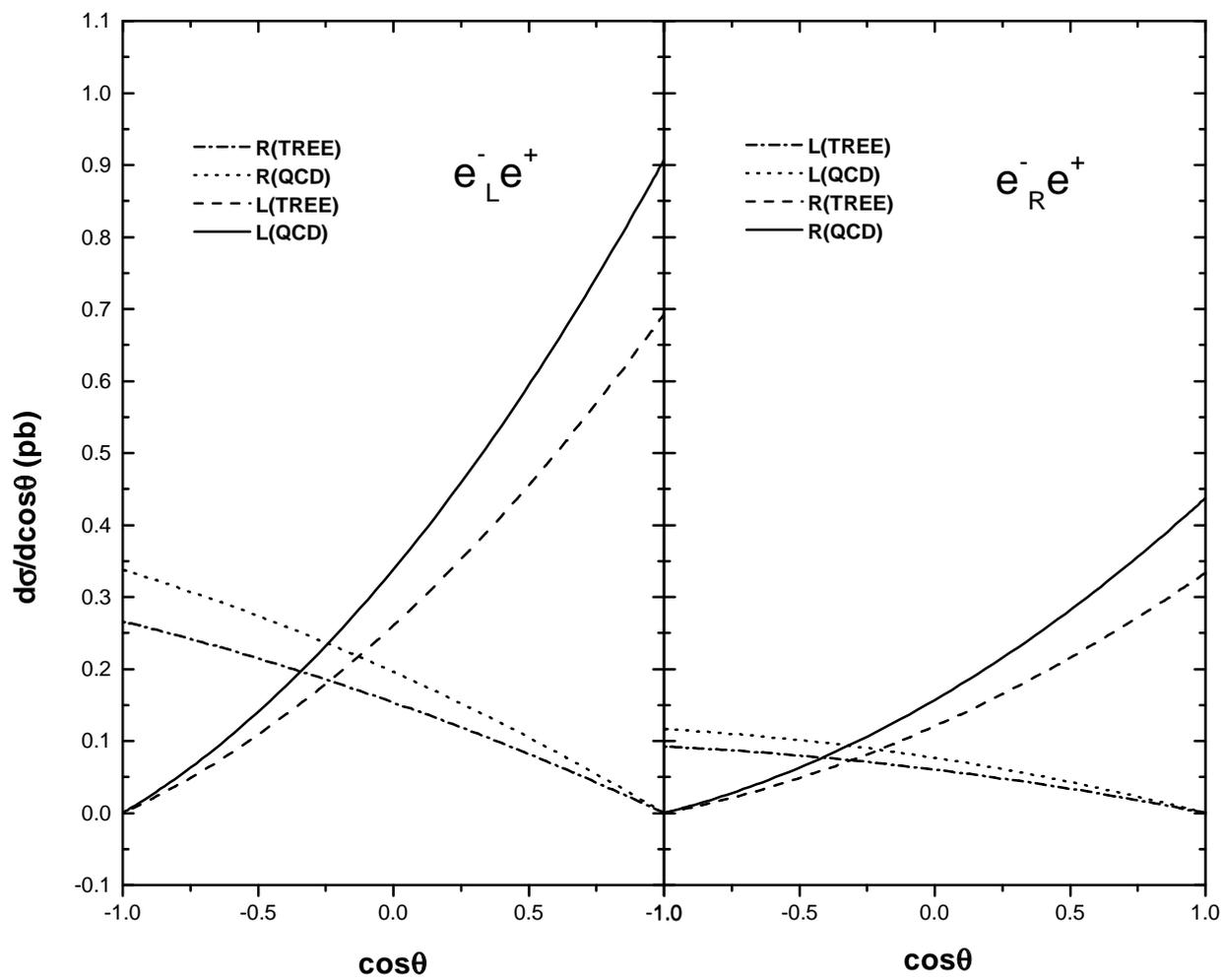
\item[Fig.4:] The differential  cross sections in the {\em Helicity
basis} for the $e_L^- e^+ \to t_{L,R} \bar{t}$ and
$e_R^- e^+ \to t_{L,R} \bar{t}$ processes at $\sqrt{s}=400 GeV$,
and assuming $\omega_g=(\sqrt{s}-2m_t)/5=10 GeV$.

\end{description}


\begin{thebibliography}{99}

\bibitem{top98}
CDF collaboration, F.Abe, {\it et. al.}, Phys.Rev.Lett. 74(1992)2626,
Phys.Rev.Lett. 80(1998)2767;
D0 collaboration, S.Abachi, {\it et. al.}, Phys.Rev.Lett. 74(1995)2632.

\bibitem{toph}
I. Bigi, H. Krasemann, Z. Phys. C7, 127(1981); J. K\"uhn, Acta.Phys.Austr.
(Suppl.) XXIV, 203 (1982);
I. Bigi, Y. Dokshitzer, V. Khoze, J.K\"uhn, and P.Zerwas,
Phys.Lett. 181B, 157(1986).

\bibitem{tt1}
C.R. Schmidt and M.E. Peskin, Phys.Rev.Lett. 69(1992)410;
D. Atwood, A. Aeppli and A. Soni, Phys.Rev.Lett. 69(1992)2754;
W. Bernreuther and A. Brandenburg, Phys.Rev. D49(1994)4481;
T. Stelzer and S. Willenbrock, Phys.Lett. B374(1996)169;
P. Haberl, O. Nachtmann and A. Wilch, Phys.Rev. D53(1996)4875;
C. Mahlon and S. Parke, Phys. Rev. D53(1996)4886,
Phys.Lett. B411(1997)173.

\bibitem{tt2}
D. Atwood and A. Soni, Phys. Rev. D45(1992)2405;
G.L. Kane, G.A. Ladinsky and C.P. Yuan, Phys.Rev. D45(1992)124;
C. P. Yuan, Phys.Rev. D45(1992)782;
G.A. Ladinsky, Phys.Rev. D46(1992)3789;
W. Bernreuther, O. Nachtmann, P.Overmann and T.Schr\"oder,
Nucl.Phys. B388(1992)53, erratum B406(1993)516;
T. Arens and L.M. Seghal, Nucl.Phys. B393(1993)46;
P. Poulose and S.D. Rindani, Phys.Lett. B349(1995)379;
S. Kuhloan, {\it et al.}, SLAC-Report-485, hep-ph/9605011;
M.M. Tung, J. Bernabeu and J. Penarrocha, Phys.Lett. B418(1998)181;
S. Accomando {\it et al.}, Phys.Rept. 299(1998)1.

\bibitem{parke96}
S. Parke and Y. Shadmi, Phys.Lett. B387(1996)199.

\bibitem{cross}
G. Grunberg, Y.J.Ng and S.-H.H. Tye, Phys.Rev. D21 (1980)62;
J. Jers\`ak, E.Laermann and P.Zerwas, Phys. Rev. D25(1982) 1218;
M.M.Tung, J.Bernab\'eu and J. Pe\"arrocha, Nucl. Phys. B470(1996)41;
S.Groote, J.G. K\"orner and J. A. Leyva, Nucl.Phys. B527(1998)3.

\bibitem{pdf98}
C. Caso {\it et. al.}, Eur.Phys. J. C3(1998)1.

\bibitem{9807209}
J. Kodaira, T.Nasuno and S.Parke, hep-ph/9807209, Phys.Rev. D59(1999)014023.

\end{thebibliography}
\end{document}